\documentclass{IEEEtran}
\usepackage{cite}
\usepackage{amsmath,amssymb,amsfonts}
\usepackage{algorithmic}
\usepackage{graphicx}
\usepackage{textcomp}
\usepackage{soul}
\usepackage{caption}
\usepackage{subcaption}
\usepackage{multirow}
\usepackage{balance}
\usepackage{threeparttable}
\usepackage{tikz, pgfplots, pgfplotstable, schemabloc}
\usetikzlibrary{shapes,shapes.geometric,arrows,fit,calc,positioning,automata,patterns}
\pgfplotsset{compat=1.11,
    /pgfplots/ybar legend/.style={
    /pgfplots/legend image code/.code={%
       \draw[##1,/tikz/.cd,yshift=-0.25em]
        (0cm,0cm) rectangle (3pt,0.8em);},
   },
}
\usepackage{booktabs}
\usepackage{rotating, graphicx}%
 \usepackage{tabularx, makecell}
\usepackage{pifont}
\newcommand{\reviews}[1]{{\textcolor{black}{#1}}}
\newcommand{\cmark}{\ding{51}}%
\newcommand{\xmark}{\ding{55}}%
\def\BibTeX{{\rm B\kern-.05em{\sc i\kern-.025em b}\kern-.08em
    T\kern-.1667em\lower.7ex\hbox{E}\kern-.125emX}}
\begin{document}

\title{A Tiny Transformer for Low-Power Arrhythmia Classification on Microcontrollers}

\author{Paola Busia,
Matteo Antonio Scrugli, Victor Jean-Baptiste Jung, Luca Benini, Paolo Meloni
\thanks{This work was supported by Key Digital Technologies Joint Undertaking (KDT JU) in EdgeAI “Edge AI Technologies for Optimised Performance Embedded Processing” project, grant agreement No 101097300.}
\thanks{Paola Busia, Matteo A. Scrugli, and Paolo Meloni are with the DIEE, University of Cagliari, Cagliari, Italy (e-mail: paola.busia@unica.it , matteoa.scrugli@unica.it , paolo.meloni@unica.it). }
\thanks{Victor J. B. Jung and Luca Benini are with the Integrated Systems Laboratory, ETH Z{\"u}rich, Z{\"u}rich, Switzerland.}
\thanks{Luca Benini is with the DEI, University of Bologna, Bologna Italy.}
\thanks{This article has been accepted for publication in IEEE Transactions on Biomedical Circuits and Systems. This is the author's version which has not been fully edited and
content may change prior to final publication. © 2024 IEEE. Personal use of this material is permitted. Permission from IEEE must be
obtained for all other uses, in any current or future media, including reprinting/republishing this material for advertising or promotional purposes, creating new
collective works, for resale or redistribution to servers or lists, or reuse of any copyrighted component of this work in other works. Please refer to the published version, available at: DOI 10.1109/TBCAS.2024.3401858}
}

\maketitle

\begin{abstract}
Wearable systems for the \reviews{continuous and real-time} monitoring of cardiovascular diseases are becoming widespread and valuable assets in diagnosis and therapy. A promising approach for real-time analysis of the electrocardiographic (ECG) signal and the detection of heart conditions, such as arrhythmia, is represented by the transformer machine learning model. Transformers are powerful models for the classification of time series, although efficient implementation in the wearable domain raises significant design challenges, to combine adequate accuracy and a suitable complexity.\par
In this work, we present a tiny transformer model for the analysis of the ECG signal, requiring only 6k parameters and reaching 98.97\% accuracy in the recognition of the 5 most common arrhythmia classes from the MIT-BIH Arrhythmia database, assessed considering 8-bit integer inference as required for efficient execution on low-power microcontroller-based devices. We explored an augmentation-based training approach for improving the robustness against electrode motion artifacts noise, resulting in a worst-case post-deployment performance assessment of 98.36\% accuracy. Suitability for wearable monitoring solutions is finally demonstrated through efficient deployment on the parallel ultra-low-power GAP9 processor, where inference execution requires 4.28ms and 0.09mJ.
\end{abstract}


\section{Introduction}\label{sec:intro}
According to the World Health Organization, cardiovascular diseases are the first cause of death worldwide~\cite{WHO}. The electrocardiographic (ECG) signal represents the commonly accepted non-invasive diagnostic and monitoring tool for the detection and recognition of anomalies in the functioning of the heart, whose early detection can help reduce their consequences and provide adequate treatment.\par
It is established that wearable solutions are of great importance for allowing continuous monitoring of chronic conditions while reducing the burden on clinicians and hospital structures~\cite{Iqbal2021}. In this context, where the privacy of the data is a relevant concern, near-sensor edge processing represents an increasingly attractive option.
Particular benefits would result from continuous monitoring for the detection of arrhythmia, which represents an anomaly of the heartbeat, showing an irregular or abnormal rhythm. \par
\reviews{Large-size transformers have recently demonstrated state-of-the-art accuracy in the recognition of different arrhythmia types~\cite{HUdeeptrans,Yan_FusingTransf}, surpassing the performance obtained with Convolutional Neural Networks (CNNs)~\cite{Zhao_efficientnet}. As a consequence, the ability to capture non-local data dependencies provided by the attention mechanism in transformers appears very promising also in the ECG processing field. Nonetheless, the complexity of current models~\cite{HUdeeptrans,Yan_FusingTransf} is not compatible with efficient execution on wearable monitoring devices: the neural network-based classification approaches presented in the literature for this processing domain are dominated by lightweight CNNs~\cite{Farag_filtersCNN,Scrugli_adaptive}, whose accuracy is still suboptimal compared to the state of the art, whereas the role of transformer models at this complexity scale still needs to be investigated.}\par 
In this work, we plan to combine the advantages deriving from the attention mechanism exploited in transformers with actual execution on wearable devices for \reviews{out-of-ambulatory continuous and real-time} monitoring. 
To this aim, we have designed a lightweight transformer architecture that can be executed on microcontrollers, \reviews{resulting from a design exploration of the topological parameters having the highest impact on its storage requirements and complexity}. Moreover, we have considered motion artifact noise resilience as a main objective within the overall design process, using signals with artifacts to augment the dataset during training.\par 
The main contributions of this work can be summarized in the following points:
\begin{itemize}
    \item we present a tiny transformer model for arrhythmia recognition, reaching accuracy aligned with the state of the art of transformer-based classifiers, with 60$\times$ fewer parameters and 300$\times$ fewer required operations;
    \item we evaluate the robustness of our proposed classifier considering common issues of real-time execution, resulting from the presence of artifacts and noise in the signal acquisition;
    \item we provide an efficient implementation on a low-power commercial MCU, resulting in less than 0.09mJ per inference obtained with quantization up to 8-bit precision and parallel execution on 8 cores, thus showing the proposed model is suitable for \reviews{continuous and real-time} wearable monitoring.
\end{itemize}
\begin{table*}[h!]
    \centering\scriptsize
            \begin{threeparttable}
    \begin{tabular}{c|c|c|c|c|c|c|c|c|c|c|c}
    \hline
        \multirow{2}{*}{\textbf{Work}} & \multirow{2}{*}{\textbf{Year}} & \multirow{2}{*}{\textbf{Classes}} & \multirow{2}{*}{\textbf{Task}} & \multirow{2}{*}{\textbf{Model}} & \multirow{2}{*}{\textbf{Accuracy}} & \multirow{2}{*}{\textbf{Memory}} & \multirow{2}{*}{\textbf{MOPS}} & \textbf{Inference} & \textbf{Energy} & \multirow{2}{*}{\textbf{Hardware Target}} & \textbf{Noise} \\
         & & & & & & & &\textbf{Time} & \textbf{Consumption} & & \textbf{Addition} \\
    \hline 
    \cite{Yan_FusingTransf} & 2019 & 4 & intra-patient & ViT & 99.62\% & 3 MB\tnote{*} & 365\tnote{*} & - & - & - &  \xmark \\
    \cite{HUdeeptrans} & 2022 & 4 & intra-patient & Transformer & 99.49\% & 20 MB & 70.53 & - & - & - &  \xmark \\
    \cite{MS_MLP} & 2022 & 5 & intra-patient & MLP & $>$ 99\% & - & 30.41 & - & - & - &  \xmark \\ 
    \cite{Zhao_efficientnet} & 2022 & 8 & intra-patient & CNN & 99.54\% &  16 MB & - & - & - &  - &  \xmark \\
        \cite{Ahmed_1dcnn} & 2023 & 4 & intra-patient & CNN & 99\% & 2.2 MB & 151 & - & - & - &  \xmark \\ 
    \cite{Scrugli_adaptive} & 2022 & 5 & intra-patient & CNN & 98.89\% & 96 kB & 0.74 & 215 ms & 0.66 mJ &  ARM Cortex-M4 &  \xmark \\ 
    \cite{Farag_filtersCNN} & 2023 & 3 & inter-patient & CNN & 98.18\% & 15 kB & 0.05 & 1 ms & - & Cortex-ARMv8 &  \cmark \\
        \hline         
    \multirow{3}{*}{\textbf{this work}} &  \multirow{3}{*}{\textbf{2024}} &  \multirow{3}{*}{\textbf{5}} &  \multirow{3}{*}{\textbf{intra-patient}} &  \textbf{ViT 32-bit} & \textbf{99.05\%} & \multirow{3}{*}{\textbf{49 kB}} & \multirow{3}{*}{\textbf{0.97}} &  \multirow{3}{*}{\textbf{4.28 ms}} & \multirow{3}{*}{\textbf{0.09 mJ}} & \multirow{3}{*}{\textbf{GAP9}} & \xmark \\
    &  &  &  & \textbf{ViT 8-bit} & \textbf{98.97\% } &  &  & &  &   &  \xmark \\
        &  &  &  &  \textbf{ViT 8-bit} & \textbf{98.36\% } &  &  &  &  & & \cmark \\
    \hline         
    \end{tabular}
           \begin{tablenotes}
\item[*] Estimated from paper.
\end{tablenotes}
    \end{threeparttable}
    \caption{\reviews{State of the art of neural network-based arrhythmia classification on the MIT-BIH Arrhythmia database, including classification performance and efficiency figures evaluated on the target hardware. We indicate with the symbol "-" that the corresponding information is not reported or not assessed in the cited paper.}}
    \label{tab:state_of_the_art}
\end{table*}
\reviews{We start with a review of the state of the art of classification approaches in Section~\ref{sec:related_works}. In Section~\ref{sec:materials}, we provide a general overview of the considered ECG monitoring system and describe its components, focusing on the architecture of the proposed transformer-based classifier.
The assessment of the classification performance is presented in
Section~\ref{sec:exp_results}. 
Finally, we report in Section~\ref{sec:deploy} the evaluation of the performance of the proposed model deployed on the GAP9 low-power multicore microcontroller unit (MCU).}\par

\section{Related Work}\label{sec:related_works}
Table~\ref{tab:state_of_the_art} summarizes the recent works presenting neural network-based models for arrhythmia recognition on the MIT-BIH dataset~\cite{MIT_BIH, Physionet}. For each of the listed works, we report in subsequent columns the publication year, the number of arrhythmia classes considered, the specific task addressed, the classification model, and the test accuracy obtained. The "\reviews{Memory", "MOPS"}, "Inference Time", and "Energy consumption" columns summarize the \reviews{complexity and} performance evaluated on the target platform, reported in the "Hardware Target" column. Finally, for each work, we indicate whether the degradation of the classification accuracy in the presence of growing levels of noise has been explored.\par
As can be noted, classification based on CNNs reaches remarkable results. In~\cite{Zhao_efficientnet} the authors present a CNN adapted from the EfficientNet family, focusing on the 8-class arrhythmia recognition task and reaching 99.54\% accuracy with around 5.3M parameters. A 1D-CNN exploiting 2M parameters is presented in~\cite{Ahmed_1dcnn}, reporting 99\% accuracy in the 4-class recognition problem. 
As an alternative, the work of~\cite{MS_MLP} presents arrhythmia classification based on a Multi-Layer-Perceptron (MLP) model, scoring a classification accuracy higher than 99.47\% for each of the 5 classes considered. \reviews{These accurate models rely on a high number of parameters, not suitable for deployment on low-power microcontrollers. To better motivate this statement, we consider as a reference for wearable deployment a maximum memory footprint of 128 kB, which is the available storage space in the L1 memory of our target platform, the GAP9 parallel ultra-low-power processor~\cite{GAP9}}.\par
Tiny CNN models have also been exploited for real-time arrhythmia detection on resource-constrained devices. Ref.~\cite{Scrugli_adaptive} presents two alternative CNNs for arrhythmia recognition on a commercial low-power microcontroller, reaching up to 98.69\% and 98.89\% accuracy in the 5-class recognition problem, with 18K and 93K parameters. The efficiency of inference execution was evaluated on the ST Sensortile microcontroller, where it requires 215ms and 0.66mJ. Hybrid models resulting from the combination of the CNN topology with the MLP, such as in~\cite{9848993Wong}, or the Long-Short-Term-Memory (LSTM) network, such as in~\cite{9669005Sivapalan}, were also considered for binary classification, to distinguish between normal and ventricular, or generally abnormal, heartbeats. The proposed models reach up to 98.5\% and 97\% accuracy. \reviews{The reduced complexity of these models results in a sub-optimal classification performance compared to the state of the art, encouraging further studies on the design of efficient models for arrhythmia recognition.}\par
\reviews{Similarly, the work of~\cite{Farag_filtersCNN}} presents a CNN working on the first derivative of the ECG signal and exploiting matched filters reproducing the different arrhythmia classes within the convolutional layers. The model also exploits the processing of the RR intervals (the distance between consecutive heartbeat peaks in the signal) through a stack of fully connected layers, obtaining 98.18\% accuracy with a 15 kB memory footprint\reviews{, addressing the more challenging inter-patient classification problem, where the classifier is trained and tested on the data from separate sets of patients.} Edge deployment on a Raspberry Pi equipped with a Cortex-ARMv8 64-bit System on Chip results in less than 1 ms inference time. \reviews{This very efficient model is thus suitable for low-power deployment, nonetheless the obtained accuracy, evaluated on only three arrhythmia classes, is still over 1\% below the state of the art.}\par 
\reviews{Another relevant contribution is presented in~\cite{10125017Liu}, introducing BioAIP, a specialized processor for biological signal processing, enabling the possibility to refine a CNN-based global classification model considering a small subset of patient-specific data, reporting up to 7.5\% accuracy improvement on tests following the leave-one-patient-out approach, reaching up to 99.16\% classification accuracy. However, this interesting work is mostly oriented to the description of the proposed processing platform, rather than describing the specifics of the proposed ECG classification and testing approach.}\par
\reviews{As can be noticed from Table~\ref{tab:state_of_the_art}, transformer-based classification represents the state-of-the-art approach, particularly based on the work} of~\cite{Yan_FusingTransf}. The authors presented a Vision Transformer (ViT) consisting of 3 encoding blocks and 4 parallel heads in the attention layer. They obtained 99.62\% accuracy, exploiting a denoising and baseline-wandering removal step, and concatenating the information about the RR intervals within the last classification layer.
An interesting approach is presented in the work of~\cite{HUdeeptrans}. The proposed model is applied on 3s-long windows of signal and provides the positioning and classification of the heartbeats occurring during that period, reaching 99.86\% heartbeat positioning accuracy, and 99.49\% classification accuracy. The best model exploits 5M parameters, but competitive results were obtained with a similar architecture based on 0.75MB of parameters, reaching 99.05\% accuracy.\par
\reviews{Other relevant examples are represented by 
the transformer architectures presented in~\cite{Che2021_linkconstr,Natarajan,MENG2022102236}, which were designed and tested on different datasets, considering multiple-lead acquisition.} \par
To the best of our knowledge, the transformer-based classifiers in the literature have storage and computational requirements non-compatible with efficient deployment on low-power MCUs. \reviews{Considering our target platform, GAP9, the model in~\cite{Yan_FusingTransf} cannot be allocated even in the larger L2 memory, which provides an additional 1.5 MB of storage space accessible with lower efficiency.}\par 
\reviews{Finally, the robustness of the classification under different noise conditions is only explored in the work of~\cite{Farag_filtersCNN}, considering white noise addition up to 3db Signal to Noise Ratio (SNR). On the contrary, we assessed the classification performance under different noise conditions, considering real recordings of electrode motion artifacts noise.}\par
In this work, we thus present an efficient transformer architecture, targeting accurate arrhythmia detection on low-power wearable devices, and extending the performance assessment to the typical challenges of real-time wearable operating conditions. As most of the referenced works, with the exception of~\cite{Farag_filtersCNN}, we refer to the intra-patient classification problem, where each heartbeat in the dataset is considered independently, and randomly assigned to the training, validation, or test set. 
\section{ECG Classifier Design}\label{sec:materials}
A typical ECG signal is shown in Figure~\ref{fig:heartwave}, referring to acquisition from a single electrode. Each heartbeat is defined by a complex wave around a peak (R) in the amplitude, commonly known as the QRS complex. The shape of the wave around the R-peak, and the intervals between consecutive peaks (pre-RR and post-RR intervals), highlighted in the figure, are both meaningful diagnostic elements.\par
\begin{figure}
 \centering
                  \includegraphics[width=\columnwidth]{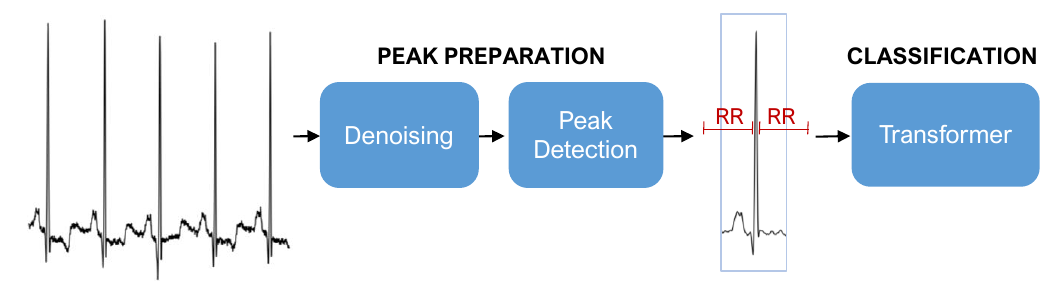} 
     \caption{Overview of the ECG monitoring system.}
     \label{fig:system_view}
\end{figure}

\input{images/heartwave}
Figure~\ref{fig:system_view} describes a general overview of the real-time monitoring system that would be executed on a wearable device. The input to the system is represented by the continuous ECG signal. 
As reported in the plot, our proposed classification system exploits two main information elements: a window of the acquired signal selected around a single heartbeat, and a pair of values representing the pre-RR and post-RR intervals. The identification and processing of these two elements is the result of the peak preparation block, which involves a denoising step and a peak detection step. The ECG window and the pre-RR and post-RR intervals are thus provided to the transformer-based classifier, for the recognition of the considered classes of arrhythmia.\par
In the following, we describe in Section~\ref{sec:preprocessing} the details of the data preparation block. The architectural description of our proposed transformer model for heartbeat classification is reported in Section~\ref{sec:transf_model}. Finally, Section~\ref{sec:dataset}
presents the main characteristics of the reference \reviews{dataset} considered for the design optimization and assessment. 
\subsection{Peak preparation}\label{sec:preprocessing}
The raw ECG signal goes through a denoising block performing baseline-wandering and noise removal. This block was structured according to the scheme suggested in~\cite{Yan_FusingTransf}, which includes two main filtering steps. The baseline wandering component is identified thanks to two median filters with respectively 200ms and 600ms window length, and subtracted from the signal. At this point, a low-pass filter with a cut-off frequency of 35 Hz is exploited for powerline noise removal.\par
The clean signal is then processed for the identification of the window of interest around the R peak. Among the common QRS detection algorithms in the literature, here we refer to the well-known Pan-Tompkins detection algorithm, which allows the detection of 99.67\% of the heartbeats in the considered dataset~\cite{PanTompkins}. 

\subsection{Transformer for Arrhythmia recognition}\label{sec:transf_model}
\begin{figure*}
 \centering
                  \includegraphics[width=\textwidth]{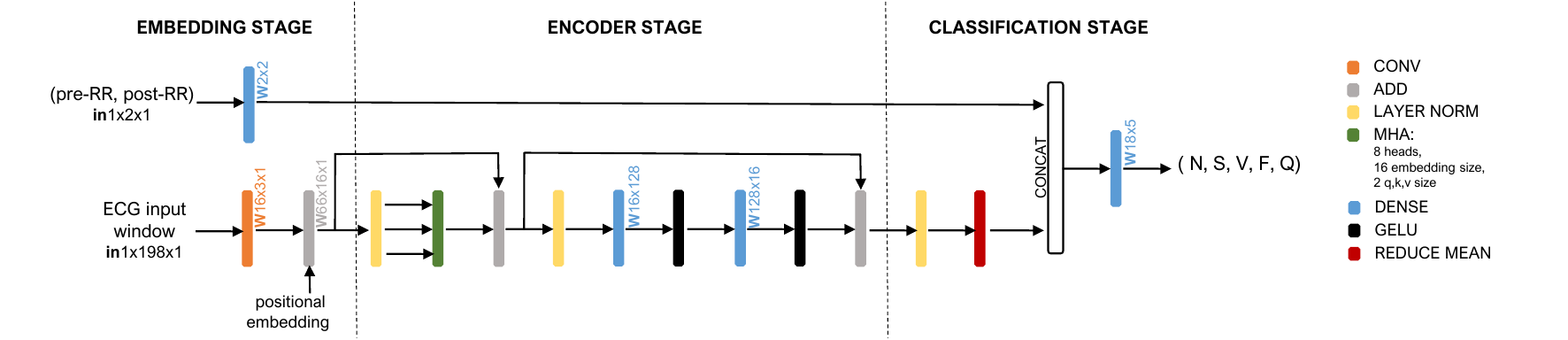} 
     \caption{\reviews{Architecture of the proposed transformer model for arrhythmia classification. The shape of the weight tensor in each layer is indicated with the prefix \textbf{W}, whereas the prefix \textbf{in} refers to the shape of the input data tensors.}}
     \label{fig:ecgformer}
\end{figure*}
The proposed classifier is a Vision Transformer~\cite{Alexey2021, Vaswani20175999}, adapted for the processing of 1D medical signals, similarly to the approaches of~\cite{Bioformers, EEGformer}\reviews{, which presented tiny vision transformer models for hand gesture recognition based on the electromyographic signal and electroencephalography processing for epilepsy monitoring}. The model proposed in this work was developed starting from the architecture presented in~\cite{EEGformer} and further optimized for the arrhythmia recognition task. 
\par
The structure of the model is represented in Figure~\ref{fig:ecgformer}.
A typical ViT model consists of three main stages, an \textit{embedding stage}, an \textit{encoder stage}, and a \textit{classification stage}. The embedding stage performs a data preparation task, producing as output a sequence of patches, inspired by the tokens of an input text, and including the positional information. 
\par
This sequence represents the input to the encoder stage, where most of the data processing is performed. The encoder is constituted by a stack of blocks including a Multi-Head-Attention (MHA) layer and a Feed Forward network, exploiting residual connections.\par 
The MHA represents the most important mechanism within the transformer, allowing the evaluation of the mutual relevance between two points within the observed time series. As a first step for the attention evaluation, the input sequence of patches is projected into three different spaces, known as queries \textit{Q}, keys \textit{K}, and values \textit{V}. The first two projections are used for the evaluation of the attention matrix, whose items allow to give weight to the elements within the third \textit{V} projection, according to the Equation in~\ref{eq:attention}, where $d$ represents the size of the \textit{Q}, \textit{K}, and \textit{V} projections.
\begin{equation}
\label{eq:attention}
Attention(\textbf{\textit{Q}},\textbf{\textit{K}},\textbf{\textit{V}})=Softmax(\frac{\textbf{\textit{QK}}^T}{\sqrt{d}})\textbf{\textit{V}}
\end{equation}
The attention mechanism is usually replicated considering multiple parallel executions, called heads, where independent projections of the input are evaluated. In MHA, a final linear projection is performed in order to restore the embedding data size. 
\reviews{Finally, the classification stage is usually implemented as an MLP or a Dense layer.}\par
\reviews{In the following, we describe the design exploration performed for the definition of the proposed transformer topology. As a starting point, we fixed a few design choices, based on the literature and previous results in~\cite{EEGformer}. In detail, in alignment with the choice in~\cite{Scrugli_adaptive}, the input size was selected equal to 198 samples, equally distributed on the left and on the right of the considered R peak position within the heartbeat, and corresponding to roughly 0.5 s of data based on the sampling frequency of the considered dataset.}\par \reviews{Additionally, we considered a general topology embedding a single encoder block, aligned with the one defined in~\cite{EEGformer}, which featured 50k parameters. Furthermore, in our model, the embedding stage is obtained with a 1D-convolutional block, as proposed in~\cite{Bioformers}. This solution adapts the Vision Transformer embedding to 1D signals. In this approach, the input signal of shape $1\times198\times1$ (\textit{channel, width, height}) is transformed into a tensor of size $E\times S\times1$, where the number of output channels $E$ is interpreted as the width of a single patch ($E\times1$), whereas $S$ defines the length of the sequence of patches to be processed and is obtained from the original $198\times1$ input based on the size $k$ of the filtering kernels (applied considering no overlap, with stride $s=k$).}\par
\reviews{Finally, according to the original topology presented in~\cite{Vaswani20175999}, we considered a fixed relation between the size of the patches processed within the attention layer - the embedding size $E$ -, the number of attention heads $H$, and the size  $P$ of the projected queries, keys, and values; the relation is set as~$P=E/H$.}\par
\reviews{The design exploration was performed to select the optimal value of the parameters impacting the computational load of the transformer, and especially of the MHA layer, namely the embedding size $E$, the length of the sequence of patches $S$, the number of heads $H$, and the size $h$ of the hidden layer in the feed-forward network executed after MHA. The exploration process aimed at maximizing the accuracy on the target dataset, under a defined storage constraint (128 kB). In this stage, the comparison of the candidate design points evaluated the classification accuracy on some left-out records, i.e. records “108”, “210”, “233”, “106”, and “203”.}\par
\reviews{The parameters were selected among a small set of alternatives, based on a sequential evaluation where all parameters were kept fixed except for the one being explored at a certain step.  We started from the embedding size, and the choice $E=32$ exploited in~\cite{EEGformer}: we evaluated values within the set \{16, 32, 48, 64\}. For this exploration step, the kernel size $k$ was set to 3, $H$ was set to 8, and $h$ was set to 128. Based on the outcome of the exploration, $E=16$ was selected and considered in the following. 
The second design choice was the kernel/stride size impacting the sequence length in MHA. We considered values within the set \{1, 3, 5, 7\}  and obtained the highest accuracy for $k = 3$, which was kept in the following. The third design choice was the number of heads $H$, selected equal to 8 among values within \{1, 2, 4, 8\}. Finally, the hidden layer size $h$ was selected equal to 128 from a set including \{32, 48, 96, 128\}.}\par 
\begin{figure*}
 \centering
\begin{subfigure}[b]{0.48\textwidth}
\begin{tikzpicture}

\definecolor{color0}{rgb}{0.83921568627451,0.152941176470588,0.156862745098039}
\definecolor{color1}{rgb}{1,0.549019607843137,0}
\definecolor{color2}{rgb}{0.133333333333333,0.545098039215686,0.133333333333333}
\definecolor{color3}{rgb}{0.254901960784314,0.411764705882353,0.882352941176471}
\definecolor{color4}{rgb}{0.117647058823529,0.564705882352941,1}

\begin{axis}[
legend cell align={left},
legend columns=5,
legend style={
  fill opacity=1,
  draw opacity=1,
  text opacity=1,
  at={(0.35,1.01)},
  anchor=south west,
  draw=white!80!black,
  font={\scriptsize},
},
height=0.6\columnwidth,
width=\columnwidth,
log basis x={10},
tick align=outside,
tick pos=left,
x grid style={white!69.0196078431373!black},
xlabel={Memory Footprint [B]},
xmin=13731.494391831, xmax=402286.465141498,
xmode=log,
xtick style={color=black},
xtick={1000,10000,100000,1000000,10000000},
xticklabels={
  \(\displaystyle {10^{3}}\),
  \(\displaystyle {10^{4}}\),
  \(\displaystyle {10^{5}}\),
  \(\displaystyle {10^{6}}\),
  \(\displaystyle {10^{7}}\)
},
y grid style={white!69.0196078431373!black},
ylabel={Accuracy \%},
ymin=73, ymax=90.5,
ytick style={color=black},
tick label style={font=\tiny},
label style={font=\small}
]
\addplot [color0, dotted, mark=*, mark size=1.75, mark options={solid}]
table {%
49396 87.85
65706 79.2
84074 76.9
104490 81.95
};
\addlegendentry{embedding size E}
\addplot [color1, dotted, mark=*, mark size=1.75, mark options={solid}]
table {%
16010 78.7
23282 83.1
49396 87.85
345034 79.6
};
\addlegendentry{kernel size k}
\addplot [color2, dotted, mark=*, mark size=1.75, mark options={solid}]
table {%
18894 75.5
23250 77.4
31962 80.03
49396 87.85
};
\addlegendentry{number of heads H}
\addplot [color3, dotted, mark=*, mark size=1.75, mark options={solid}]
table {%
46122 85.1
46666 80.03
47210 81.8
48298 76.76
49396 87.85
};
\addlegendentry{hidden size h}
\addplot [white!66.2745098039216!black]
table {%
131072 65
131072 90.5
};
\path [draw=white!66.2745098039216!black, fill=white!66.2745098039216!black, opacity=0.3]
(axis cs:131072,73)
--(axis cs:131072,90.5)
--(axis cs:402286.465141498,90.5)
--(axis cs:402286.465141498,73)
--cycle;
\addlegendentry{128 kB constraint}
\addplot [black, mark=triangle*, mark size=4, mark options={solid}]
table {%
49396 87.85
};
\draw (axis cs:65682,79.185) node[
  scale=0.7,
  anchor=base west,
  text=color0,
  rotate=0.0
]{E 32};
\draw (axis cs:84044,76.885) node[
  scale=0.7,
  anchor=base west,
  text=color0,
  rotate=0.0
]{E 48};
\draw (axis cs:104454,81.935) node[
  scale=0.7,
  anchor=base west,
  text=color0,
  rotate=0.0
]{E 64};
\draw (axis cs:15000,77.6) node[
  scale=0.7,
  anchor=base west,
  text=color1,
  rotate=0.0
]{k 7};
\draw (axis cs:18000,83) node[
  scale=0.7,
  anchor=base west,
  text=color1,
  rotate=0.0
]{k 5};
\draw (axis cs:300000,78.6) node[
  scale=0.7,
  anchor=base west,
  text=color1,
  rotate=0.0
]{k 1};
\draw (axis cs:18900,75) node[
  scale=0.7,
  anchor=base west,
  text=color2,
  rotate=0.0
]{H 1};
\draw (axis cs:23258,77) node[
  scale=0.7,
  anchor=base west,
  text=color2,
  rotate=0.0
]{H 2};
\draw (axis cs:31967,79.6) node[
  scale=0.7,
  anchor=base west,
  text=color2,
  rotate=0.0
]{H 4};
\draw (axis cs:48000,85) node[
  scale=0.7,
  anchor=base west,
  text=color4,
  rotate=0.0
]{h 32};
\draw (axis cs:47000,80.022) node[
  scale=0.7,
  anchor=base west,
  text=color4,
  rotate=0.0
]{h 48};
\draw (axis cs:47220,81.792) node[
  scale=0.7,
  anchor=base west,
  text=color4,
  rotate=0.0
]{h 64};
\draw (axis cs:48308,76.752) node[
  scale=0.7,
  anchor=base west,
  text=color4,
  rotate=0.0
]{h 96};
\draw (axis cs:40000, 88.7) node[
  scale=0.7,
  anchor=base west,
  text=black,
  rotate=0.0
]{E16 k3 H8 h128};
\end{axis}
\end{tikzpicture}
\subcaption{}
\label{fig:exploration_Arch_a}
\end{subfigure}
\begin{subfigure}[b]{0.48\textwidth}
    \begin{tikzpicture}

\definecolor{color0}{rgb}{0.83921568627451,0.152941176470588,0.156862745098039}
\definecolor{color1}{rgb}{1,0.549019607843137,0}
\definecolor{color2}{rgb}{0.133333333333333,0.545098039215686,0.133333333333333}
\definecolor{color3}{rgb}{0.254901960784314,0.411764705882353,0.882352941176471}
\definecolor{color4}{rgb}{0.117647058823529,0.564705882352941,1}

\begin{axis}[
legend cell align={left},
legend style={
  fill opacity=0.8,
  draw opacity=1,
  text opacity=1,
  at={(0.03,0.03)},
  anchor=south west,
  draw=white!80!black,
  font = \scriptsize
},
height=0.607\columnwidth,
width=\columnwidth,
tick align=outside,
tick pos=left,
x grid style={white!69.0196078431373!black},
xlabel={\# Parameters},
xmin=3000, xmax=40152.2,
log basis x={10},
xmode=log,
xtick={2000,5000,10000,20000,40000},
xticklabels={
  \(\displaystyle {2*10^{3}}\),
  \(\displaystyle {5*10^{3}}\),
  \(\displaystyle {10^{4}}\),
  \(\displaystyle {2*10^{4}}\),
  \(\displaystyle {4*10^{4}}\),
},
xtick style={color=black},
y grid style={white!69.0196078431373!black},
ylabel={Accuracy \%},
ymin=73, ymax=90.5,
ytick style={color=black},
tick label style={font=\tiny},
label style={font=\small}
]
\addplot [color0, dotted, mark=*, mark size=1.75, mark options={solid}]
table {%
6639 87.85
15173 79.2
25765 76.9
38405 81.95
};
\addplot [color1, dotted, mark=*, mark size=1.75, mark options={solid}]
table {%
6085 78.7
6229 83.1
6639 87.85
8709 79.6
};
\addplot [color2, dotted, mark=*, mark size=1.75, mark options={solid}]
table {%
6629 75.5
6629 77.4
6629 80.03
6639 87.85
};
\addplot [color3, dotted, mark=*, mark size=1.75, mark options={solid}]
table {%
3461 85.1
3989 80.03
4517 81.8
5573 76.76
6639 87.85
};
\addplot [black, mark=triangle*, mark size=4, mark options={solid}]
table {%
6639 87.85
};
\draw (axis cs:15149,79.185) node[
  scale=0.7,
  anchor=base west,
  text=color0,
  rotate=0.0
]{E 32};
\draw (axis cs:25735,76.55) node[
  scale=0.7,
  anchor=base west,
  text=color0,
  rotate=0.0
]{E 48};
\draw (axis cs:30000,81.935) node[
  scale=0.7,
  anchor=base west,
  text=color0,
  rotate=0.0
]{E 64};
\draw (axis cs:6070,78.705) node[
  scale=0.7,
  anchor=base west,
  text=color1,
  rotate=0.0
]{k 7};
\draw (axis cs:5200,83.105) node[
  scale=0.7,
  anchor=base west,
  text=color1,
  rotate=0.0
]{k 5};
\draw (axis cs:8694,79.605) node[
  scale=0.7,
  anchor=base west,
  text=color1,
  rotate=0.0
]{k 1};
\draw (axis cs:6634,75.5) node[
  scale=0.7,
  anchor=base west,
  text=color2,
  rotate=0.0
]{H 1};
\draw (axis cs:6634,77.4) node[
  scale=0.7,
  anchor=base west,
  text=color2,
  rotate=0.0
]{H 2};
\draw (axis cs:6634,80.03) node[
  scale=0.7,
  anchor=base west,
  text=color2,
  rotate=0.0
]{H 4};
\draw (axis cs:3471,85.092) node[
  scale=0.7,
  anchor=base west,
  text=color4,
  rotate=0.0
]{h 32};
\draw (axis cs:3800,78.5) node[
  scale=0.7,
  anchor=base west,
  text=color4,
  rotate=0.0
]{h 48};
\draw (axis cs:4527,81.792) node[
  scale=0.7,
  anchor=base west,
  text=color4,
  rotate=0.0
]{h 64};
\draw (axis cs:5200,75.5) node[
  scale=0.7,
  anchor=base west,
  text=color4,
  rotate=0.0
]{h 96};
\draw (axis cs:6639, 88.7) node[
  scale=0.7,
  anchor=base west,
  text=black,
  rotate=0.0
]{E16 k3 H8 h128};
\end{axis}
\end{tikzpicture}
\subcaption{}
\label{fig:exploration_Arch_b}
\end{subfigure}
     \caption{\reviews{Exploration of the architectural parameters based on left-out test accuracy on the MIT-BIH dataset. Each dot in the plot represents one of the evaluated design points, resulting from the selection of the parameters explored: embedding size, kernel size, number of heads, and hidden layer size. Dot placement in a) considers the memory footprint of the model, accounting for the storage of the partial results produced during inference computation, whereas placement in b) considers the number of parameters in the model. The selected point is highlighted in black. Shaded grey area: outside of the 128 kB feasibility region.} }
     \label{fig:exploration_Arch}
\end{figure*}
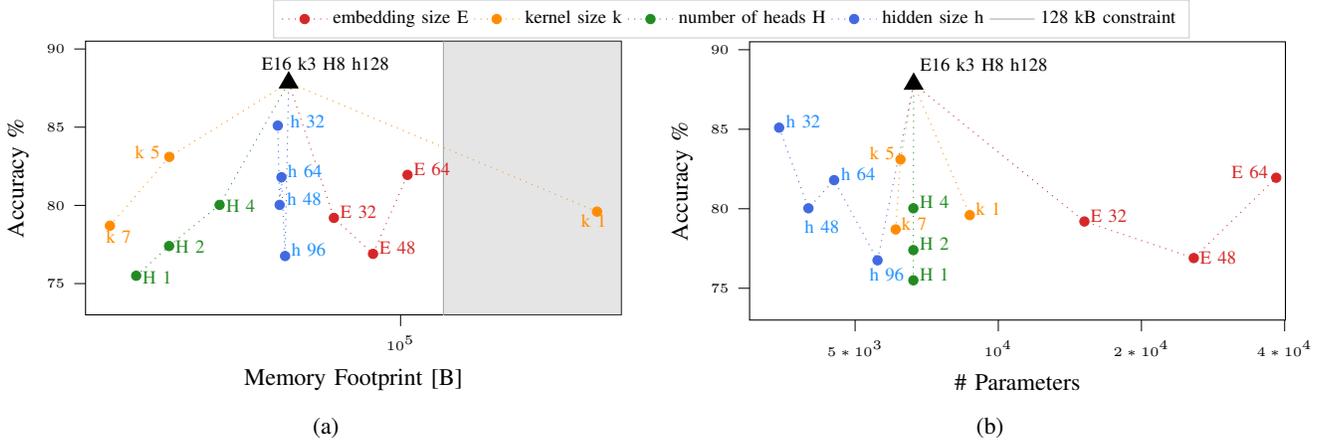
\reviews{The outcome of the exploration is summarized in Figures~\ref{fig:exploration_Arch_a} and~\ref{fig:exploration_Arch_b}, where each evaluated point is placed according to its accuracy and memory footprint or parameters' count. The comparison between the two plots highlights how the memory footprint is highly impacted by the storage space allocated for the partially computed results, especially the attention matrix. As a consequence, design choices having a limited impact on the parameters' count, such as the kernel size in the embedding convolutional layer, have a very significant effect on the overall memory requirements of the model. As can be noticed, all the evaluated design points fit the memory constraint defined for the exploration, except for the solution corresponding to $k = 1, E = 16, H= 8$, and $h = 128$. The selected point, resulting in the highest accuracy, is highlighted in the plot.}\par
\reviews{The selected model architecture is summarized in Table~\ref{tab:ecgformer_params}, where the first column highlights the composition of each stage, the second column lists the operands applied  during the execution of the model, the third column reports the shape of the output data produced by the computation, the fourth column summarizes the values of the parameters selected based on the design exploration, and the fifth one details the parameters' count for each operand.}\par
\reviews{Considering the results reported in other works in the literature~\cite{Farag_filtersCNN, Wang_CNN_entropy, HUdeeptrans, Yan_FusingTransf}, presenting neural network models for arrhythmia  recognition and demonstrating the benefits of including the information about the RR intervals to differentiate the considered arrhythmia classes, we also included this information as an additional input to the transformer model}. We thus followed the example of~\cite{Yan_FusingTransf} and introduced an additional input to the traditional transformer topology, representing the distance between the current R peak and the previous and following one. This second input is first linearly projected within the embedding stage and then concatenated to the output of the encoder which processes the ECG window around the heartbeat.\par
The tuple of values representing the RR intervals is normalized into the range [-2,2], to resemble the distribution of the
encoder output at the concatenation level, thus simplifying the quantization process.

\begin{table}
    \centering\scriptsize
    \begin{tabular}{c|c|c|c|c}
        \hline
        \multirow{2}{*}{\textbf{Stage}} & \multirow{2}{*}{\textbf{Layer}}  & \multirow{2}{*}{\textbf{Output}} & \multicolumn{2}{c}{\textbf{Parameters}} \\
        & & & \textbf{Values} & \textbf{\#} \\
        \hline
        \multirow{4}{*}{\textit{Embedding}} & \multirow{2}{*}{Convolution} &  \multirow{2}{*}{16$\times$66$\times$1} &k: 3$\times$1 & \multirow{2}{*}{64}\\
        &  & & s: 3 & \\
        \cline{2-5}
        & Add Pos. Embedding  & 16$\times$66$\times$1 & & 1056 \\
        \cline{2-5}
        & Dense  & 1$\times$2 & & 4\\
                \hline
        \multirow{10}{*}{\textit{Encoder}} & Layer Norm &   16$\times$66$\times$1 & & 32 \\
        \cline{2-5}
        & \multirow{4}{*}{Multi-Head-Attention}  & \multirow{4}{*}{16$\times$66$\times$1} &  E: 16 & \multirow{4}{*}{1088} \\
        & &  & S: 66 & \\
        &  & & d: 2 & \\
        &  & & H: 8 & \\
        \cline{2-5}
        & Layer Norm & 16$\times$66$\times$1 & & 32 \\
        \cline{2-5}
        & Dense & 128$\times$66$\times$1 & & 2176 \\
        \cline{2-5}
        & Gelu & 128$\times$66$\times$1 &  \\
        \cline{2-5}
        & Dense  & 16$\times$66$\times$1 & & 2064 \\
        \cline{2-5}
        & Gelu  & 16$\times$66$\times$1 &  \\
                        \hline
        \multirow{3}{*}{\textit{Classification}} & Layer Norm &   16$\times$66$\times$1 & & 32 \\
        \cline{2-5}
        & Reduce Mean  & 1$\times$16 & \\
        \cline{2-5}
        & Dense  & 1$\times$5 & & 95\\
        \hline
\textit{Tot} & & & & 6643 \\
        \hline
    \end{tabular}
            \caption{\reviews{Parameters of the proposed transformer model for arrhythmia classification.}}
    \label{tab:ecgformer_params}
\end{table}

\subsection{\reviews{Reference Dataset}}\label{sec:dataset}
As a main reference for this study, we considered the MIT-BIH Arrhythmia Database~\cite{MIT_BIH, Physionet}, which collects the ECG recordings of 47 subjects into 48 records, lasting 30 minutes. The data included in the dataset is a composition of randomly selected ambulatory recordings from Boston's Beth Israel Hospital, and recordings selected to include the clinically relevant arrhythmia examples. Each record reports the signal acquired by two channels, usually modified limb lead II (MLII) and V1, with a 360Hz sampling frequency. We refer in the following to processing based on the signal acquired by the MLII.\par
In alignment with most of the works in the literature~\cite{HUdeeptrans, Yan_FusingTransf, Farag_filtersCNN, Wang_CNN_entropy}, we excluded from the training and test sets the records containing paced beats, namely "102", "104", "107", "217". Furthermore, we referred to the AAMI standard~\cite{association1999testing} to identify the 5 most relevant classification groups for the arrhythmia recognition problem: N (non-ectopic beats), S (supra-ventricular ectopic beats), V (ventricular ectopic beats), F (fusion beat), Q (unclassifiable beat).\par 
Real-time heartbeat classification on wearable monitoring devices can be affected by different sources of noise. 
\reviews{The MIT-BIH Noise Stress Test Database~\cite{MITBIHnoise,Physionet} includes authentic ECG recordings accompanied by real noise samples, essential for simulating typical disturbances encountered in ambulatory ECG recordings. This dataset comprises twelve half-hour ECG recordings and three additional half-hour segments explicitly containing noise sourced from three primary disturbances: baseline wander, muscle artifacts, and electrode motion artifacts.
The noise records were derived from recordings using standard ECG equipment and physically active volunteers, with electrodes positioned on the limbs to avoid interfering with the ECG signal visibility. This setup effectively replicates the challenging conditions often faced in real-world ECG data collection.
The focus of this research is particularly on the noise caused by electrode motion artifacts, the most problematic type due to its potential to mimic ectopic beats and its resistance to simple filtering techniques. This real noise environment is critical for analyzing the robustness of ECG signal processing techniques under realistic conditions.}\par
Here we describe the process leading to the creation of the noise-corrupted \reviews{data} referenced in the Experimental Section~\ref{sec:add_noise}. The corruption of the signal is indicated as different Signal-to-Noise ratio (SNR) levels, expressed in dB. For brevity, we indicate as noiseless the original signal, with no additional noise deriving from electrode motion artifacts. 
Given a noiseless signal \( S \) from the MIT-BIH Arrhythmia Database and a noise component \( N \) from the MIT-BIH Noise Stress Test Database, the noise was scaled by a scaling factor \( \alpha \) calculated based on the desired SNR, according to Equation~\ref{eq:snr2}. 

\begin{equation}
\alpha = \sqrt{ \frac{ \text{mean}(|S|^2) } { \text{mean}(|N|^2) \times 10^{ \frac{\text{desired SNR}}{10} } } }
\label{eq:snr2}
\end{equation}

The resulting noisy signal \( Y \) is generated in Equation \ref{eq:snr1}, where the scaling factor \( \alpha \) results from Equation \ref{eq:snr2}.

\begin{equation}
Y = S + \alpha \times N
\label{eq:snr1}
\end{equation}

Each noisy ECG signal is annotated with the original arrhythmia labels from the MIT-BIH Arrhythmia Database.\par
\reviews{The MIT-BIH dataset was chosen as a reference for the performance assessment due to the extensive literature targeting this resource. However, the ECG classification system we are proposing can be exploited for arrhythmia recognition based on single-lead acquisition, provided that proper upsampling/downsampling of the data is performed to obtain a coherent input shape corresponding to 0.5 s observation interval around the detected R peak.}

\subsection{Evaluation Metrics}\label{sec:metrics}
As will be further detailed, Table~\ref{tab:dataset_split_composition} reports the typical composition of the training and test sets from the MIT-BIH dataset. As can be noticed, the dataset is highly unbalanced, presenting a majority of normal heartbeats and a different number of instances for each of the represented classes. Due to this reason, for the performance assessment, we not only considered the overall classification accuracy but also the sensitivity and precision on each of the targeted arrhythmia classes. The definition we referenced for these metrics is reported in equations~\ref{eq:sensitivity}
 and~\ref{eq:precision}, where we apply them to class N:

\begin{equation}\label{eq:sensitivity}
    Sens_N = \frac{TN_N}{TN_N + FS_N + FV_N + FF_N + FQ_N}
\end{equation}

\begin{equation}\label{eq:precision}
    Prec_N = \frac{TN_N}{TN_N + FN_S + FN_V + FN_F + FN_Q}
\end{equation}

where the notation $TN_N$ indicates a normal sample, correctly classified as normal, $FS_N$ indicates a normal sample which is classified as a supra-ventricular ectopic beat, and finally $FN_S$ indicates a supra-ventricular ectopic beat classified as normal. A similar logic can be applied to the interpretation of the other symbols in the equations.

\section{Experimental Results}\label{sec:exp_results}
In the following, we summarize the outcome of the tests performed to evaluate the performance of our proposed arrhythmia classifier.

\subsection{Intra-patient arrhythmia recognition}\label{sec:intra-patient_test}

We assessed the performance of our proposed model in the intra-patient classification task, based on 5-fold cross-validation. The ECG records in the dataset were segmented around each heartbeat corresponding to one of the NSVFQ arrhythmia classes, considering the dataset annotations for the heartbeat positioning. All collected segments were shuffled and randomly split into a training set, a validation set, and a test set, with a 7:1:2 ratio, to enable a direct comparison with the literature~\cite{Yan_FusingTransf}. The composition of the dataset and a typical train-valid-test split is reported in Table~\ref{tab:dataset_split_composition}.\par

\begin{table}[]
    \centering\scriptsize
    \begin{tabular}{c|c|c|c|c}
    \hline
        \textbf{Arrhytmia Class} & \textbf{\# Samples} & \textbf{Train} & \textbf{Valid} & \textbf{Test} \\
        \hline
         N & 90098 & 63022 & 9057 & 18019 \\ 
         S & 2781 & 1967 & 273 & 541 \\ 
         V & 7007 & 4919 & 664 & 1424 \\ 
         F & 802 & 573 & 75 & 154 \\ 
         Q & 15 & 12 & 1 & 2 \\ 
         \hline
    \end{tabular}
    \caption{Composition of the MIT-BIH dataset and typical training - validation - test split considered.}
    \label{tab:dataset_split_composition}
\end{table}

The training of the classifier was performed on one NVIDIA T4 Tensor Core GPU, using the TensorFlow framework inside the Google Colab environment. We exploited 200 epochs training, with Adam optimizer, adaptive reduce-on-plateau learning rate with starting value $2e-3$, and batch size 128.\par
The classification performance for full-precision inference is reported in Table~\ref{tab:assessment_xclass}, in terms of mean value and standard deviation across the 5 tests considered. The performance of the model is only slightly impacted by the extremely low sensitivity and precision registered for the Q class, which is very under-represented in the dataset, and excluded from most studies~\cite{Yan_FusingTransf, HUdeeptrans}. Our proposed model reaches an average of 99.05\% accuracy on the 5 splits considered. Table~\ref{tab:confusion_matrix} reports the summary confusion matrix referring to the cross-validation test. As can be derived from Table~\ref{tab:assessment_xclass}, the main weakness is represented by a reduced sensitivity in the recognition of the F class.\par

\begin{table}[]
    \centering\scriptsize
 \begin{tabular}{cl|r|r|r|r|r}
 \parbox[t]{2mm}{\multirow{5}{*}{\rotatebox[origin=c]{90}{\textbf{True Labels}}}} & \textbf{N} & 89867 & 119 & 79 & 31 & 2\\
  \cline{2-7}
 & \textbf{S} & 289 & 2468 & 23 & 1 & 0\\
  \cline{2-7}
 & \textbf{V} & 139 & 31 & 6783 & 45 & 0\\
  \cline{2-7}
 & \textbf{F} & 120 & 4 & 62 & 616 & 0\\
 \cline{2-7}
 & \textbf{Q} & 10 & 0 & 4 & 0 & 1\\
 \cline{3-7}
  & \multicolumn{1}{c}{} & \multicolumn{1}{c|}{\textbf{N}} & \multicolumn{1}{c|}{\textbf{S}} & \multicolumn{1}{c|}{\textbf{V}} & \multicolumn{1}{c|}{\textbf{F}} & \multicolumn{1}{c}{\textbf{Q}} \\
    & \multicolumn{1}{c}{} & \multicolumn{5}{c}{\textbf{Predicted Labels}} \\
 \end{tabular}
            \caption{Confusion matrix summarizing performance on the considered splits for the full-precision model, when no additional noise is applied to the signal.}
    \label{tab:confusion_matrix}
 \centering
        \begin{tabular}{c|c|c}
    \hline
    \textbf{Class} & \textbf{Sensitivity} & \textbf{Precision}  \\
    \hline
    N  & 99.74\% (0.08) & 99.38\% (0.04)  \\
    S  & 88.79\% (2.02) & 94.14\% (1.72) \\
    V  & 96.91\% (0.58) & 97.59\% (0.24)\\
    F  & 76.92\% (5.48) & 88.99\% (1.93) \\
    Q  & 10\% (20)  & 10\% (20) \\
    \hline
    \textbf{Accuracy tot} & \multicolumn{2}{c}{\textbf{99.05\% (0.08)}} \\
    \hline
    \end{tabular}
    \caption{Test classification performance, when no additional noise is applied to the signal. We report mean values and standard deviation in 5-fold cross-validation with full-precision inference.}
  \label{tab:assessment_xclass}
 \end{table}
     \par

\subsection{Ablation Study}\label{sec:ablation}
We report in the following the ablation study of the different data pre-processing choices explored, summarized in Table~\ref{tab:ablation}, limiting the analysis to a single train-test split.\par The first line in the Table reports the accuracy of the selected model, depicted in Figure~\ref{fig:ecgformer} of Section~\ref{sec:transf_model}. 
We started from the removal of the concatenation of the second input, representing the RR interval information, which helps in the arrhythmia class discrimination, as reported in several previous studies~\cite{Farag_filtersCNN, Wang_CNN_entropy, HUdeeptrans, Yan_FusingTransf}. The result was a 0.36\% accuracy drop, reported in the second line.\par
We assessed at this point the relevance of the denoising step. We report in the third line the performance obtained when considering the RR information, but the selected topology was trained, validated, and tested on unfiltered data, not pre-processed with the denoising step in Figure~\ref{fig:system_view}. The resulting accuracy is lower than the best obtained by 0.51\% points.
To confirm this result, we finally removed both the RR concatenation and the denoising step, thus obtaining the performance reported in the fourth line. As can be observed, there is no further accuracy degradation compared to the number reported in the third line, showing how the RR information only provides a relevant advantage when combined with an effective denoising approach. 
\begin{table}[]
    \centering
    \begin{tabular}{c|c}
    \hline
     \textbf{Model} & \textbf{Test Accuracy}  \\
         \hline
        Selected Model & 99.17\% \\
        no RR & 98.81\% \\
        no denoising & 98.66\% \\
        no RR and no denoising & 98.69\% \\
             \hline

    \end{tabular}
    \caption{Comparison of test performance on a single 7:1:2 split, based on different input processing choices. Accuracy numbers refer to tests performed with no additional noise applied to the signal.}
    \label{tab:ablation}
\end{table}

\subsection{Post-deployment evaluation}\label{sec:add_noise}
This work targets wearable devices for real-time heartbeat classification, we thus present in this section the assessment of the performance degradation deriving from two main challenges of wearable deployment, namely increasing levels of noise corrupting the ECG signal and accuracy loss due to data quantization.\par

\begin{figure}
    \centering

\begin{tikzpicture}

\definecolor{crimson2143940}{RGB}{214,39,40}
\definecolor{darkgray176}{RGB}{176,176,176}
\definecolor{forestgreen4416044}{RGB}{44,160,44}
\definecolor{gray}{RGB}{128,128,128}
\definecolor{lightgray}{RGB}{211,211,211}
\definecolor{lightgray204}{RGB}{204,204,204}
\definecolor{orange}{RGB}{255,165,0}
\definecolor{royalblue}{RGB}{65,105,225}

\begin{axis}[
legend cell align={left},
legend style={
  fill opacity=0.1,
  draw opacity=1,
  text opacity=1,
  at={(0.03,0.03)},
  anchor=south west,
  draw=none,
  font = \scriptsize
},
height = 0.7\columnwidth,
width = \columnwidth,
tick align=outside,
tick pos=left,
x grid style={darkgray176},
xlabel={Test condition},
xmin=0.8, xmax=5.2,
xtick style={color=black},
xtick={1,2,3,4,5},
xticklabels={noiseless,24dB SNR,10dB SNR,3dB SNR,balanced mix},
y grid style={lightgray,dotted},
ylabel={Accuracy \%},
ymajorgrids,
ymin=88, ymax=99.575,
ytick style={color=black},
tick label style={font=\scriptsize},
xlabel style = {font=\scriptsize},
ylabel style = {font=\scriptsize},
]
\addplot [thick, gray, mark=*, mark size=1, mark options={solid}]
table {%
1 99.17
2 98.95
3 96
4 91.3
5 96.36
};
\addlegendentry{training with noiseless signal}
\addplot [orange, mark=*, mark size=1, mark options={solid,fill=white}]
table {%
1 99.02
2 98.98
3 95.92
4 91.07
5 96.25
};
\addlegendentry{training with 24dB SNR}
\addplot [royalblue, mark=*, mark size=1, mark options={solid,fill=white}]
table {%
1 98.75
2 98.78
3 98.67
4 97.43
5 98.41
};
\addlegendentry{training with 10dB SNR}
\addplot [forestgreen4416044, mark=*, mark size=1, mark options={solid,fill=white}]
table {%
1 98.9
2 98.86
3 98.66
4 98.21
5 98.67
};
\addlegendentry{training with 3dB SNR}
\addplot [thick, crimson2143940, mark=*, mark size=1, mark options={solid,fill=white}]
table {%
1 99.15
2 99.15
3 98.79
4 97.98
5 98.77
};
\addlegendentry{training with balanced mix}
\end{axis}

\end{tikzpicture}

    \caption{Test classification performance considering increasing levels of noise deriving from electrode motion artifacts.}
    \label{fig:noise_accuracydrop}
\end{figure}
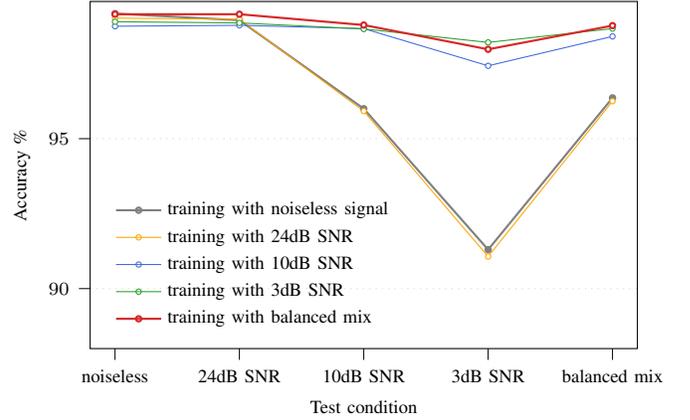

\subsubsection{Noise degradation} As a first step for the noise degradation analysis, we used as a baseline the accuracy reported in  Table~\ref{tab:ablation}, deriving from the evaluation on the noiseless signal, and the corresponding specific train-test split.
We repeated the same test using noise-augmented versions of the dataset samples in training and test, to create Figure~\ref{fig:noise_accuracydrop}. Different lines correspond to different levels of noise used for augmenting the training set, while different evaluation points in each line correspond to a different level of noise in the test. Considered levels are \textit{noiseless}, \textit{24dB SNR}, \textit{10dB SNR}, \textit{3dB SNR}, and \textit{balanced mix}, where \textit{noiseless} corresponds to the original non-augmented dataset samples and \textit{balanced mix} corresponds to an equally partitioned test set, with different noise levels applied to each subset, to emulate real-life conditions where the noise level is expected to vary over an evaluation.\par 
The solid grey line shows the results obtained when performing classification on different test conditions with 
the model trained on the non-corrupted signal. As can be observed, in this case, the accuracy is dramatically affected by the noise, with up to almost 8\% points drop registered with high noise levels and test SNR equal to 3dB.\par
In order to improve robustness during real-time inference, we considered the effect of including examples of corrupted signals during the training process, represented by the other lines in the plot. As can be observed, augmenting with a specific level of noise improves the resilience of the classification algorithm when a similar noise level is considered in the test, although in general, it determines a slight degradation for lower noise levels. 
The \textit{balanced mix} strategy, on the other hand, performs better than specific levels in most evaluation points and especially in the leftmost one, representing real-life conditions. It reduces the accuracy drop on the most corrupted signal (SNR = 3dB) to less than 1.2\% points. This gain in the robustness to high noise levels does not compromise the classification accuracy on the noiseless signal, which is only 0.02\% lower than the highest one reported in Table~\ref{tab:ablation}. 
The results thus confirmed the advantages of a training approach considering a variable noise level for augmentation.\par
Based on these findings, we repeated 5-fold cross-validation using training data augmentation and reported the results in Table~\ref{tab:summary_augmonclean}. The first two columns summarize the performance for the classification of the noiseless signal. The average accuracy is 99.07\%, thus no degradation is observed compared to the un-augmented training approach. The confusion matrix is also included for completeness and reported in Table~\ref{tab:confusion_matrix_augmonclean}.\par
\begin{table}[]
    \centering\scriptsize
    \begin{tabular}{c|c|c|c|c}
    \cline{2-5}
    & \multicolumn{2}{c|}{\textbf{Noiseless Test}} & \multicolumn{2}{c}{\textbf{Balanced Mix Test}} \\
    \hline
    \textbf{Class} & \textbf{Sensitivity} & \textbf{Precision} & \textbf{Sensitivity} & \textbf{Precision}  \\
    \hline
    N  & 99.73\% (0.08) & 99.42\% (0.08)  & 99.59\% (0.12) & 99.15\% (0.08) \\
    S  & 89.13\% (1.1) & 93.64\% (1.34) & 84.08\% (1.2) & 91.34\% (1.62)\\
    V  & 97\% (0.42) & 98.01\% (0.33)  & 95.7\% (0.42) & 96.59\% (0.47)\\
    F  & 78.73\% (4.97) & 88.63\% (3.42) & 71.05\% (4.34) & 84.62\% (5.15)\\
    Q  & 4\% (8)  & 10\% (20)  & 2\% (4)  & 2.5\% (5) \\
    \hline
    \textbf{Accuracy} & \multicolumn{2}{c|}{\textbf{99.07\% (0.06)}} & \multicolumn{2}{c}{\textbf{98.65\% (0.07)}} \\
    \hline
    \end{tabular}
    \caption{Test classification performance, reporting mean values and standard deviation in 5-fold cross-validation with full-precision inference, when training data augmentation with varying noise levels is exploited.}
    \label{tab:summary_augmonclean}

 \centering
 \begin{tabular}{cl|r|r|r|r|r}
 \parbox[t]{2mm}{\multirow{5}{*}{\rotatebox[origin=c]{90}{\textbf{True Labels}}}} & \textbf{N} & 89858 & 136 & 50 & 37 & 17\\
  \cline{2-7}
 & \textbf{S} & 278 & 2479 & 22 & 2 & 0\\
  \cline{2-7}
 & \textbf{V} & 130 & 33 & 6789 & 44 & 2\\
  \cline{2-7}
 & \textbf{F} & 104 & 1 & 64 & 631 & 2\\
 \cline{2-7}
 & \textbf{Q} & 11 & 0 & 3 & 0 & 1\\
 \cline{3-7}
  & \multicolumn{1}{c}{} & \multicolumn{1}{c|}{\textbf{N}} & \multicolumn{1}{c|}{\textbf{S}} & \multicolumn{1}{c|}{\textbf{V}} & \multicolumn{1}{c|}{\textbf{F}} & \multicolumn{1}{c}{\textbf{Q}} \\
    & \multicolumn{1}{c}{} & \multicolumn{5}{c}{\textbf{Predicted Labels}} \\
 \end{tabular}
              \caption{Confusion matrix summarizing \textit{noiseless} test performance on the considered splits for the full-precision model, when training data augmentation with varying noise levels is exploited.}
    \label{tab:confusion_matrix_augmonclean}
 \centering
 \begin{tabular}{cl|r|r|r|r|r}
 \parbox[t]{2mm}{\multirow{5}{*}{\rotatebox[origin=c]{90}{\textbf{True Labels}}}} & \textbf{N} & 358901 & 710 & 483 & 232 & 66\\
  \cline{2-7}
 & \textbf{S} & 1630 & 9355 & 130 & 8 & 1\\
  \cline{2-7}
 & \textbf{V} & 807 & 182 & 26802 & 192 & 9\\
  \cline{2-7}
 & \textbf{F} & 596 & 4 & 325 & 2278 & 5\\
 \cline{2-7}
 & \textbf{Q} & 47 & 0 & 11 & 0 & 2\\
 \cline{3-7}
  & \multicolumn{1}{c}{} & \multicolumn{1}{c|}{\textbf{N}} & \multicolumn{1}{c|}{\textbf{S}} & \multicolumn{1}{c|}{\textbf{V}} & \multicolumn{1}{c|}{\textbf{F}} & \multicolumn{1}{c}{\textbf{Q}} \\
    & \multicolumn{1}{c}{} & \multicolumn{5}{c}{\textbf{Predicted Labels}} \\
 \end{tabular}
              \caption{Confusion matrix summarizing \textit{balanced mix} test performance on the considered splits for the full-precision model, when training data augmentation with varying noise levels is exploited.}
    \label{tab:confusion_matrix_noisy}
 \end{table}
On the other hand, a significant advantage was obtained on the \textit{balanced mix} test. In this scenario, 98.65\% was achieved on average on the 5 train-test splits considered. The corresponding cumulative confusion matrix is reported in Table~\ref{tab:confusion_matrix_noisy}. As can be observed, each test sample was replicated in order to represent a different noise condition in real-time inference. The resulting evaluation metrics are summarized in the right columns of Table~\ref{tab:summary_augmonclean}. The comparison with the performance on the \textit{noiseless} test shows a generally reduced sensitivity and precision, especially in the recognition of the S and F classes.

\subsubsection{Quantization}\label{sec:quantization}
Finally, in order to optimize the storage requirements of our model and exploit at best the byte processing resources on edge-processing platforms, we made use of the Quantlab framework~\cite{Spallanzani2022_tinyMLSummit_QuantLab} to obtain 8-bit quantization. \reviews{To allow for efficient integer-only execution, the quantization strategy applies the solutions proposed in~\cite{kim2021bert} for the implementation of non-linear operators, especially Layer Norm, which is replaced with an iterative approximation, exploiting 32-bit integer parameters.}\par 
\reviews{We considered for quantization} the models trained with noise augmentation and whose full-precision performance is summarized in Table~\ref{tab:summary_augmonclean}. We exploited 15 epochs of quantization-aware fine-tuning, obtaining an average test accuracy of 98.97\% on the clean noiseless signal, with a limited 0.1\% drop from the accuracy of the full-precision models. The performance metrics are summarized in the first two columns of Table~\ref{tab:summary_8bitclean}, whereas the confusion matrix is reported in Table~\ref{tab:confusion_matrix_8bitclean}. The accuracy drop resulted from restricting all computations to the integer type, and it mostly affects the sensitivity of the S class, as can be observed from the table.\par
We finally tested the performance on the classification of noise-corrupted signal, with the \textit{balanced mix} test. The outcome is reported in the third and fourth columns of Table~\ref{tab:summary_8bitclean}, and a summary confusion matrix is reported in Table~\ref{tab:confusion_matrix_8bitnoisy}. The assessment resulted in an average 98.36\% accuracy, which we can consider as a worst-case post-deployment real-time performance prediction.\par

\begin{table}[]
    \centering\scriptsize
    \begin{tabular}{c|c|c|c|c}
    \cline{2-5}
    & \multicolumn{2}{c|}{\textbf{Noiseless Test}} & \multicolumn{2}{c}{\textbf{Balanced Mix Test}} \\
    \hline
    \textbf{Class} & \textbf{Sensitivity} & \textbf{Precision} & \textbf{Sensitivity} & \textbf{Precision}  \\
    \hline
    N  & 99.74\% (0.05) & 99.3\% (0.08) & 99.41\% (0.14) & 99.01\% (0.1)  \\
    S  & 84.73\% (1.1) & 94.65\% (0.87) & 78.3\% (1.55) & 92.12\% (0.8) \\
    V  & 97.52\% (0.42) & 97.24\% (0.45) & 96.34\% (0.53) & 94.11\% (1.26) \\
    F  & 76.58\% (3.52) & 89.56\% (2.57) & 68.94\% (2.62) & 81.33\% (4.45) \\
    Q  & 9\% (11.14)  & 40\% (48.99) & 5.75\% (7.57) & 4.48\% (6.84)\\
    \hline
    \textbf{Accuracy} & \multicolumn{2}{c|}{\textbf{98.97\% (0.04)}} & \multicolumn{2}{c}{\textbf{98.36\% (0.08)}} \\
    \hline
    \end{tabular}
    \caption{Test classification performance, reporting mean values and standard deviation in 5-fold cross-validation with 8-bit inference, when training data augmentation with varying noise levels is exploited.}
    \label{tab:summary_8bitclean}

 \centering
 \begin{tabular}{cl|r|r|r|r|r}
 \parbox[t]{2mm}{\multirow{5}{*}{\rotatebox[origin=c]{90}{\textbf{True Labels}}}} & \textbf{N} & 89860 & 115 & 88 & 32 & 3\\
  \cline{2-7}
 & \textbf{S} & 388 & 2356 & 37 & 0 & 0\\
  \cline{2-7}
 & \textbf{V} & 115 & 17 & 6825 & 40 & 1\\
  \cline{2-7}
 & \textbf{F} & 121 & 1 & 66 & 614 & 0\\
 \cline{2-7}
 & \textbf{Q} & 10 & 0 & 3 & 0 & 2\\
 \cline{3-7}
  & \multicolumn{1}{c}{} & \multicolumn{1}{c|}{\textbf{N}} & \multicolumn{1}{c|}{\textbf{S}} & \multicolumn{1}{c|}{\textbf{V}} & \multicolumn{1}{c|}{\textbf{F}} & \multicolumn{1}{c}{\textbf{Q}} \\
    & \multicolumn{1}{c}{} & \multicolumn{5}{c}{\textbf{Predicted Labels}} \\
 \end{tabular}
              \caption{Confusion matrix summarizing \textit{noiseless} test performance on the considered splits for the 8-bit model, when training data augmentation with varying noise levels is exploited.}
    \label{tab:confusion_matrix_8bitclean}
     \centering
 \begin{tabular}{cl|r|r|r|r|r}
 \parbox[t]{2mm}{\multirow{5}{*}{\rotatebox[origin=c]{90}{\textbf{True Labels}}}} & \textbf{N} & 358267 & 647 & 1061 & 344 & 73\\
  \cline{2-7}
 & \textbf{S} & 2126 & 8708 & 282 & 7 & 1\\
  \cline{2-7}
 & \textbf{V} & 743 & 93 & 26971 & 169 & 16\\
  \cline{2-7}
 & \textbf{F} & 645 & 3 & 345 & 2211 & 4\\
 \cline{2-7}
 & \textbf{Q} & 46 & 0 & 9 & 0 & 5\\
 \cline{3-7}
  & \multicolumn{1}{c}{} & \multicolumn{1}{c|}{\textbf{N}} & \multicolumn{1}{c|}{\textbf{S}} & \multicolumn{1}{c|}{\textbf{V}} & \multicolumn{1}{c|}{\textbf{F}} & \multicolumn{1}{c}{\textbf{Q}} \\
    & \multicolumn{1}{c}{} & \multicolumn{5}{c}{\textbf{Predicted Labels}} \\
 \end{tabular}
              \caption{Confusion matrix summarizing \textit{balanced mix} test performance on the considered splits for the 8-bit model, when training data augmentation with varying noise levels is exploited.}
    \label{tab:confusion_matrix_8bitnoisy}
 \end{table}

\input{images/Pie_cycles}
\subsection{Comparison with the state of the art}\label{sec:discussion}
We discuss at this point how our proposed solution compares to the state-of-the-art context, considering as a main reference the work of~\cite{Yan_FusingTransf}, which represents the most accurate alternative. Table~\ref{tab:soa_compare} summarizes the most relevant performance figures for the comparison. The first line refers to full-precision inference performed on noiseless data, as we indicate the data not corrupted with electrode motion noise. A limited accuracy drop is observed when evaluating the performance of the 8-bit model, in the second line.  
Our result presents only a slight degradation with respect to the performance of the state-of-the-art transformer model, corresponding to only a 0.65\% accuracy drop, despite the reduced complexity in terms of the number of parameters and memory footprint, reported in the fifth and sixth lines.\par
Finally, although a direct comparison with the reference model is not possible, we provide an assessment of worst-case performance considering a deployment scenario corrupted by different levels of noise. The training data augmentation allowed us to limit the accuracy drop to 0.4\% points for full-precision inference, whereas post-deployment assessment of the quantized model results in 98.36\% average accuracy, which we report in Table~\ref{tab:state_of_the_art} as the worst-case expected real-time deployment performance.

\begin{table}[]
    \centering
    \begin{threeparttable}
    \begin{tabular}{c|c|c}
    \hline
      \textbf{Metric} & \textbf{This Work}  & \cite{Yan_FusingTransf}   \\
      \hline 
      32-bit Accuracy Noiseless & 99.05\%  & 99.62\%\\
      8-bit Accuracy Noiseless & 98.97\% & - \\
      32-bit Accuracy Noisy & 98.65\% & - \\
      8-bit Accuracy Noisy & 98.36\% & - \\
      Parameters & 6649 & 405711 \tnote{1} \\
      \reviews{Memory} & 49kB &  3MB\tnote{1} \\ 
      \hline   
    \end{tabular}
    \begin{tablenotes}
     \item[1] Estimated from the paper.
     \end{tablenotes}

    \end{threeparttable}
             \caption{Comparison with the state-of-the-art transformer model for arrhythmia recognition.}
    \label{tab:soa_compare}
\end{table}
\section{Deployment}\label{sec:deploy}
In this section, we assess the efficiency of the proposed transformer classifier, considering its deployment on the parallel ultra-low power (PULP) GAP9 processor~\cite{GAP9}. This commercial platform embeds a computing cluster of 8 parallel processors, accessing a shared 128kB L1 memory to enable efficient parallel execution. The offload of computations to the cluster is handled by an additional core, working as a Fabric Controller. The memory system includes also a 1.5MB L2 memory and supports voltage and frequency tuning for power management. Based on the recent assessment on the tiny-ML benchmarks~\cite{MLcommons}, GAP9 shows exceptional energy efficiency, with as low as 0.33mW/GOP, thus representing a perfect fit for \reviews{continuous and real-time} wearable monitoring tasks. Furthermore, the computational workload of the transformer model is intrinsically parallel, thus the computing cluster can be efficiently exploited to reduce inference time during real-time execution.\par
For efficient deployment, we considered the 8-bit model reported in Table~\ref{tab:soa_compare}.
We exploited the Dory code generation tool~\cite{burrello2020dory} for assisted implementation on the target platform. As depicted in Figure~\ref{fig:ops_workload_a}, over 67\% of the computing time is occupied by the MHA layer, whose implementation exploits the parallel resources on the platform by balancing the computation among the cores of the computing cluster: each head is mapped into one of the eight cores, according to the strategy described in~\cite{microcontroller9524173}. \reviews{As can be noticed from the plots in Figures~\ref{fig:ops_workload_b} and~\ref{fig:ops_workload_c}, the MHA layer is still the most relevant workload when reducing the number of parallel heads to 4, and it is comparable to the dense layers when further reducing it to 1. To evaluate the impact of this design parameter on the overall efficiency, based on our parallelization strategy, we report in Figure~\ref{fig:ops_workload_d} the achievable speedup in terms of  required execution time and required number of OPS when considering parallelization on the 8 cores of the processing platform. As can be noticed, our implementation choices result in limited performance gains when reducing the number of heads below the number of available processing cores, despite a linear reduction of the workload.}\par 
\reviews{Table~\ref{tab:deploy} summarizes the performance figures evaluated at the working frequencies resulting in the lowest required inference time or energy consumption. At the maximum working frequency supported by the platform, 370 MHz, inference requires as low as 2.85 ms. On the other hand, the most energy-efficient configuration, obtained at the maximum frequency supported with low voltage supply, i.e. 240 MHz, results in 4.28 ms/inference time and 0.09 mJ energy consumption, for parallel execution exploiting 8 cores.}

\reviews{The execution of signal denoising and peak detection on the platform requires 0.77 ms for each newly acquired sample. System level latency accounts for 0.55 s, required for the acquisition of the window of 198 samples around the peak, plus a varying interval until the next peak detection, plus 0.77 ms to perform denoising on the last acquired sample, plus 4.28 ms inference time.}\par
The power measurements were performed with a Power Profiler Kit II (PPK2) connected to the GAP9 Evaluation Kit, resulting in an average power consumption of 20.33mW. 
\reviews{As a final consideration on the efficiency of the proposed transformer-based system, the comparison with the hardware performance metrics, reported in Table~\ref{tab:state_of_the_art} for the referenced state-of-the-art works, demonstrates a reduced energy consumption per inference, despite a slightly increased computational complexity compared to the works of~\cite{Farag_filtersCNN} and~\cite{Scrugli_adaptive}. Nonetheless, this results in improved classification accuracy, obtained with a reduced number of trainable parameters compared to the model in~\cite{Scrugli_adaptive}.}
\begin{table}
    \centering\scriptsize
    \begin{tabular}{c|c|c}
    & \multicolumn{2}{c}{\textbf{GAP9}} \\
    \hline
    \textbf{Frequency} & 370 MHz & 240 MHz \\
    \textbf{Time/Inf} & 2.85 ms & 4.28 ms \\
    \textbf{Power} & 42.60 mW & 20.33 mW \\ 
    \textbf{Energy/Inf} & 0.12 mJ & 0.09 mJ \\
             \hline
    \end{tabular}
            \caption{On-hardware performance on the GAP9 processor for parallel execution on the 8 cores of the computing cluster.}
    \label{tab:deploy}
\end{table}

\section{Conclusions}
In this work, we presented an efficient transformer model for arrhythmia classification, reaching 99.05\% accuracy in the classification of the 5 most common arrhythmia classes from the MIT-BIH Arrhythmia database. The classification performance was assessed considering different inference conditions, resembling real-time disturbance, by introducing different levels of noise corruption deriving from electrode motion artifacts, based on the noise samples from the MIT-BIH Noise Stress Test database.\par
Integer 8-bit inference resulted in 98.97\% accuracy on the noiseless signal, which is only 0.65\% lower than the state-of-the-art transformer model, although reducing by 60$\times$ the number of parameters, and by 300$\times$ the number of operations. The lean topology of the proposed model can be efficiently deployed on low-power devices for wearable monitoring, as demonstrated by the performance evaluated on the GAP9 parallel processor, where inference is executed in 4.28ms, with 0.09mJ energy consumption.
\reviews{The future directions for the development of this research include considering different implementation targets, aiming at improving the response time and energy efficiency of the proposed ECG processing system.}

\bibliographystyle{IEEEtran}  
\bibliography{references} 

\begin{IEEEbiography}[
{
\includegraphics[width=1in,height=1.25in,clip,keepaspectratio]{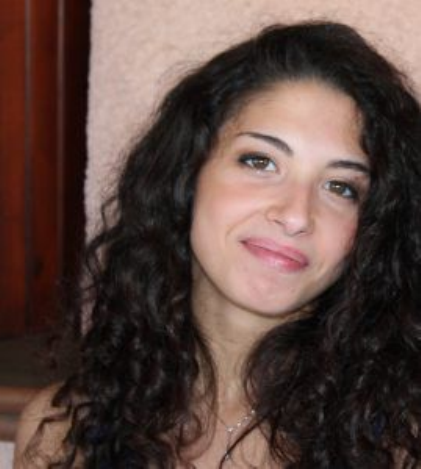}
}
]
{Paola Busia} received the Ph.D. degree in Electronics and Computer Engineering from Università degli studi di Cagliari, Italy, in 2023. She is currently assistant professor at Università degli studi di Cagliari. Her main research interests involve power efficient processing architectures for embedded applications and for at-the-edge AI.
\end{IEEEbiography}
\vskip -20 pt plus -1fil
\begin{IEEEbiography}[
{
\includegraphics[width=1in,height=1.25in,clip,keepaspectratio]{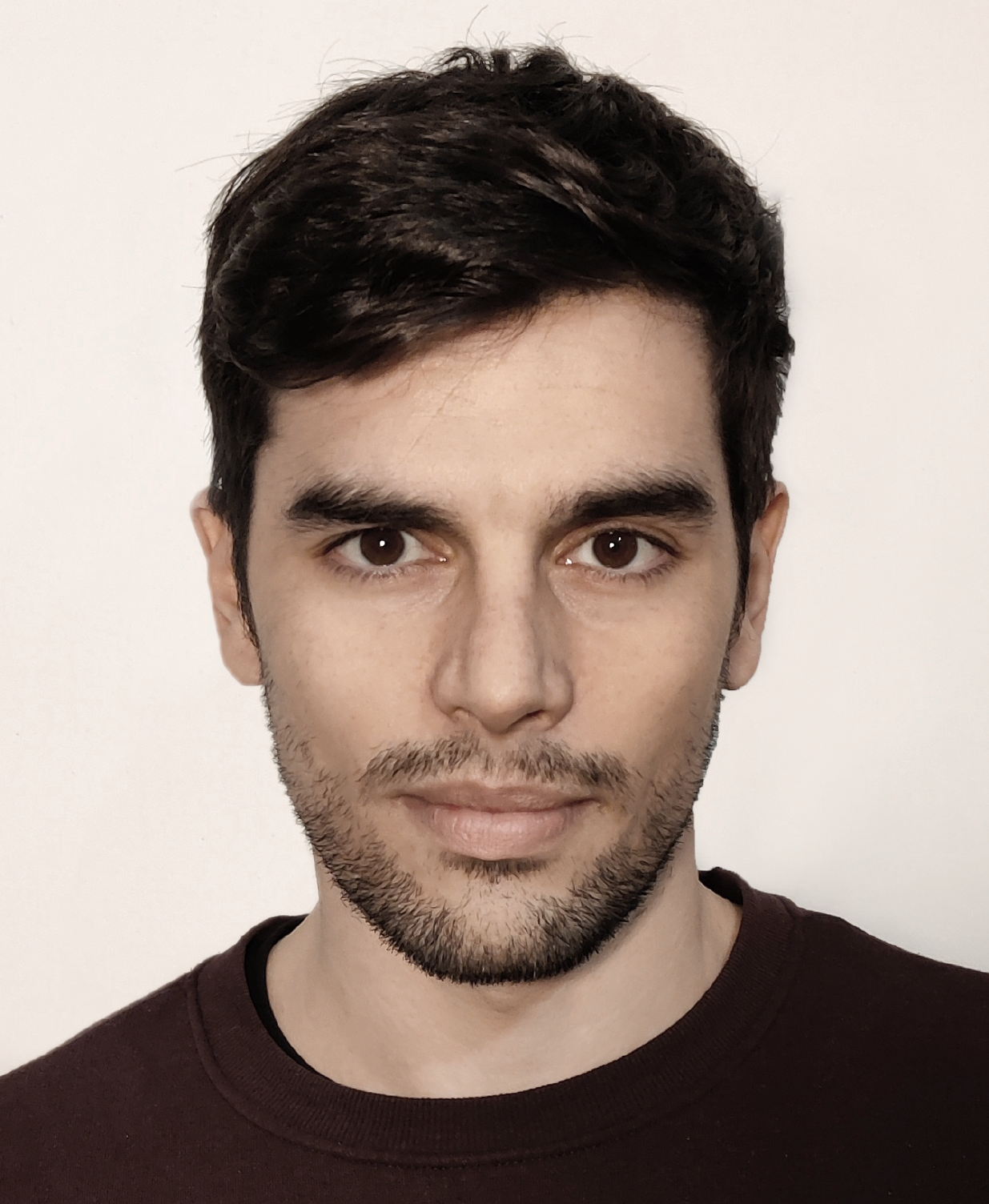}
}
]
{Matteo A.Scrugli} received his M. S. degree with honors in 2018 from the University of Cagliari and his Ph.D. in 2022 in Electrical Engineering from the University of Cagliari, where he is currently a research fellow at the DIEE. His research mainly concerns the development of systems capable of managing at runtime the hardware and software configuration of low-power devices in order to adapt it to the required operating mode. His activity is currently focused on cognitive IoT devices, based on single-core or multi-core platforms.
\end{IEEEbiography}
\vskip -20 pt plus -1fil
\begin{IEEEbiography}[
{
\includegraphics[width=1in,height=1.25in,clip,keepaspectratio]{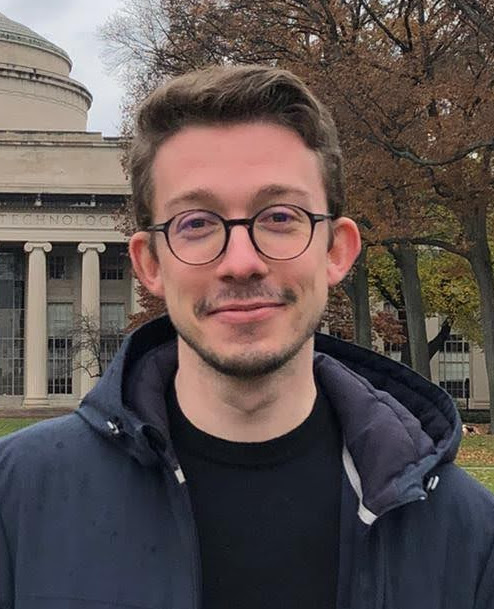}
}
]
{Victor Jean-Baptiste Jung} received his Bachelor's Degree in Computer Science and Engineering Physics from Juniata College, and the Master's Degree in Computer Science from the Institut Supérieur de l’Electronique et du Numérique of Lille (ISEN Lille) in 2022. After 3 months as a research intern with KU Leuven’s MICAS Research group supervised by Prof. Marian Verhelst, he started his Ph.D. at the Integrated Systems Laboratory with Prof. Dr. Luca Benini. His current research interests include Efficient deployment of ML models on Microcontrollers, Tiny Transformers, Scheduling and Quantization.
\end{IEEEbiography}
\vskip -20 pt plus -1fil
\begin{IEEEbiography}[
{
\includegraphics[width=1in,height=1.25in,clip,keepaspectratio]{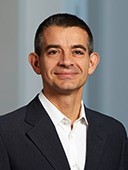}
}
]
{Luca Benini} received the Ph.D. degree in electrical engineering from Stanford University, Stanford, CA, USA, in 1997. He is the Chair of digital circuits and systems with ETH Zürich, Zürich, Switzerland, and a Full Professor with the University of Bologna, Bologna, Italy. His research interests are in energy-efficient computing, machine learning hardware and smart microsystems. Prof. Benini is a member of the Academia Europaea and a Fellow of the ACM and the IEEE.
\end{IEEEbiography}
\vskip -20 pt plus -1fil
\begin{IEEEbiography}[
{
\includegraphics[width=1in,height=1.25in,clip,keepaspectratio]{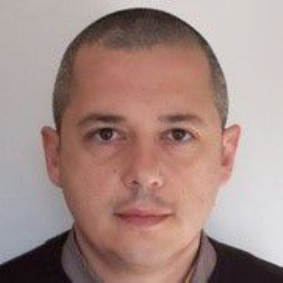}
}
]
{Paolo Meloni} is associate professor at University of Cagliari. His research activity is on the development of advanced digital systems, on the application-driven design and programming of multi-core on-chip architectures and FPGAs. \par
\end{IEEEbiography}
\balance

\end{document}